\def   \ni {\noindent}

\def   \ssk {\vskip  5truept}

\def   \bsk {\vskip 15truept}
 
\def   \newpage {\vfill\eject}
\def   \newline {\hfil\break}

\documentstyle[epsfig]{article}
\begin{document}

\hsize 5truein
\vsize 8truein
\font\abstract=cmr8
\font\keywords=cmr8
\font\caption=cmr8
\font\references=cmr8
\font\text=cmr10
\font\affiliation=cmssi10
\font\author=cmss10
\font\mc=cmss8
\font\title=cmssbx10 scaled\magstep2
\font\alcit=cmti7 scaled\magstephalf
\font\alcin=cmr6 
\font\ita=cmti8
\font\mma=cmr8
\def\ref{\par\noindent\hangindent 15pt}
\null


\title{\ni AGN MODELS FOR THE X and $\gamma$--ray BACKGROUNDS}                                               

\bsk \bsk
\author{\ni Andrea Comastri}                                                       
\bsk
\affiliation{Osservatorio Astronomico di Bologna, via Ranzani 1, 
I--40127 Bologna, Italy}                                                
\bsk
\baselineskip = 12pt

\abstract{ABSTRACT \ni

The origin of the X--ray background spectral intensity has been a long 
standing problem in high energy astrophysics research.
Deep X--ray surveys carried out with ROSAT and ASCA combined with the
broad band spectral results of Ginga and BeppoSAX satellites 
strongly support the 
hypothesis that the bulk of the X--ray background is due to the integrated 
contribution of discrete sources (mainly AGNs). 
At higher energies the unexpected findings of the Compton Gamma Ray
Observatory indicate that also the $\gamma$--ray background is likely
to be due to AGNs. 
I will discuss AGN--based models for the high energy backgrounds 
and how future observations will improve our understanding 
of the X-- and $\gamma$--ray backgrounds and of the physics and evolution
of AGNs.
}                                                    
\ssk
\baselineskip = 12pt

\ssk
\baselineskip = 12pt


\text {\ni 1. INTRODUCTION
\ssk
\ni     

The high energy extragalactic background emission covers an extremely
wide range from about 1 keV up to 100 GeV. A compilation of the 
most recent observational results is shown in Fig. 1.
For the purpose of the present discussion it is convenient to take 300 keV
as the dividing energy between the X--ray and $\gamma$--ray backgrounds
(hereinafter XRB and GRB respectively).
The shape of the XRB spectrum in the 3--60 keV energy range
is well fitted (Gruber 1992)
by a thermal model with an e--folding energy of $\sim$ 40 keV :
I(E) = 7.877 $\cdot$ E$^{-0.29} \cdot$ exp(--E/41.13 keV) 
keV cm$^{-2}$ s$^{-1}$ sr$^{-1}$ keV$^{-1}$. 
At higher energies (60--300 keV) 
the XRB slope can be approximated with a power law 
F(E) $\propto$ E$^{-\alpha}$ of energy index $\alpha \simeq$ 1.6, 
while in the 2--20 keV region a flat $\alpha \simeq$ 0.4 slope is 
required. 

Above a few hundreds of keV the measurements of the extragalactic
GRB are made difficult by the presence of high instrumental backgrounds. 
Before the launch of the Compton Gamma Ray Observatory (CGRO) several 
balloon experiments indicated an excess of flux at a few MeV (the so called
``MeV bump") with respect to the extrapolation of the XRB to higher energies.
The first all sky survey in the 0.8--30 MeV energy range carried 
out by the COMPTEL detector onboard CGRO have clearly demonstrated 
that the ``MeV bump" does not exist anymore (see Weidenspointner et al. 1999
for a compilation of recent results). An independent estimate in the
0.3--7 MeV energy range obtained by SMM (Watanabe et al. 1997) is in 
agreement with the COMPTEL results.
Even if the present measurements are affected by
rather large uncertainities, 
the high energy background spectrum seems to smoothly connect with the 
extrapolation of the XRB. For photon energies greater than a few tens 
of MeV in addition to the instrumental background one has 
to properly subtract the Galactic
diffuse emission which originates from the cosmic


\newpage 

\ni
rays interaction with the interstellar matter. 
The increased sensitivity and low instrumental 
background of the EGRET detector onboard CGRO allowed to extend and improve
the GRB discovery observation performed with SAS--2 (Fichtel et al. 1978).  
The most recent EGRET measurements of the extragalactic GRB 
(Sreekumar et al. 1998) are well represented by a single power law
with $\alpha$ = 1.10 $\pm$ 0.03 from 30 MeV to 120 GeV. The low energy 
extrapolation well matches the high energy COMPTEL data. 

\begin{figure}
\centerline{\psfig{file=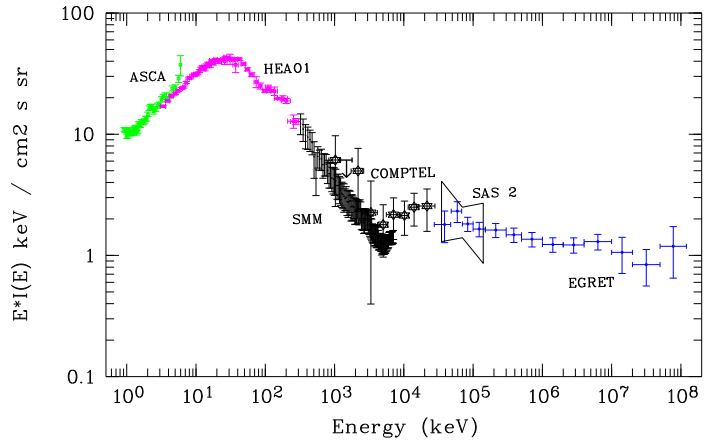, width=10cm, angle=-90}}
\caption{FIGURE 1. The extragalactic background spectrum from a compilation
of observations by several instruments (labeled). 
}
\end{figure}


The energy density of the XRB and GRB integrated from 1 keV to 100 GeV is 
$w_{X,\gamma} \simeq 7 \times 10^{-5}$ eV cm$^{-3}$.
From figure 1 it is clearly seen that the peak of the total extragalactic 
emission occurs in the X--ray domain 
at $\sim$ 30 keV. 
As a result the XRB carries a large fraction (about 80~\%) of the 
total energy density. 
The origin of the extragalactic high energy emission and in particular of 
the XRB has been, almost since its discovery, one of the main  
problems in high energy astrophysics. After several years of 
research efforts, the achievements obtained by recent satellite  
observations combined with theoretical models have significantly improved 
our knowledge, strongly suggesting that 
discrete sources provide the most important contribution 
to the whole extragalactic background spectral intensity.

\ssk
\ni 2. AGN MODELS FOR THE XRB   
\ssk
\ni 

The absence of any detectable distortion in the CMB blackbody spectrum 
as measured by COBE (Wright et al. 1994) implies that the contribution of a hot 
($ > 10^8$ K) intergalactic gas, originally suggested to explain
the XRB Bremsstrahlung--like shape, can not exceed 10$^{-5}$ of the observed
emission, leaving discrete sources as the only viable candidates.
Which is the nature of the X--ray sources making the XRB ?
Recent ultra--deep X--ray surveys carried out with ROSAT PSPC and HRI 
detectors in the 0.5--2.0 keV band have already resolved a large
fraction (70--80~\%) of the soft XRB (Hasinger et al. 1998). Optical
follow--up spectroscopy indicates that the majority of the sources
are broad emission line AGNs (Schmidt et al. 1998).
Medium--deep X--ray surveys carried 
out with ASCA (Cagnoni, Della Ceca, Maccacaro 1998; Ueda et al. 1998) 
and BeppoSAX (Giommi et al. 1998)
have started to resolve a sizeable fraction 
($\sim$ 30--40~\%) of the 2--10 keV XRB. First optical
identifications suggest that also at higher energies AGNs constitute
the dominant population (Boyle et al. 1998; Fiore et al. 1998).
The main difficulty of modeling the XRB with the integrated emission of
AGNs has been that their hard 2--20 keV X--ray spectra
are much steeper ($\alpha \simeq$ 0.7--1.0) than that of the XRB 
(the so--called spectral paradox).
It has been recognized already ten years ago (Setti \& Woltjer 1989) that 
the ``paradox" can be solved 
assuming an important contribution from sources with spectral shapes 
flattened by absorption. 
In the framework of popular AGN unified schemes
the XRB spectral intensity can be reproduced by the combined emission of 
unobscured (type 1) and obscured (type 2) 
Seyfert galaxies and quasars  with a range 
of column densities and luminosities
(Matt \& Fabian 1994; Madau et al. 1994; Comastri et al. 1995, hereinafter C95). 
All these models rely on several assumptions on the sources spectral shapes,
X--ray luminosity function (XLF) and cosmological evolution.
Given the large number of ``free parameters" it is not surprising that the
XRB spectral intensity can be well reproduced even for rather different
choices of the above described parameters. 
An attempt to reduce the parameter space has been made by 
C95 who developed a self-consistent approach 
which simultaneously fits the XRB spectrum and others available 
constraints: namely the source counts, redshift and absorption distributions 
observed in the soft (0.5--2 keV) and hard (2--10 keV) bands.
In this model the only ``free" parameter,
which is varied until a global good fit is obtained (C95, Fig. 4),
is the absorption distribution of type 2 objects, while the intrinsic spectrum
and the evolution of the XLF, parameterized as pure luminosity evolution, 
are those of type 1. 
This implies the existence of a large population of high luminosity 
highly obscured objects called type 2 QSOs.

The optical identifications of faint X--ray sources discovered in 
ROSAT, ASCA and BeppoSAX surveys has opened the possibility to 
test AGN synthesis models. Not surprisingly it seems that this rather 
simple basic model needs to be updated in order to account for the recent 
results. In particular a new estimate of the evolution of the soft 
XLF from ROSAT data indicates a more complex behaviour
than previously thought, which is best parameterized by luminosity 
dependent density evolution (Miyaji, Hasinger \& Schmidt 1999).
Moreover, the lack of type 2 QSO
in medium--deep X--ray surveys has been considered as a prove against
their existence (Halpern et al. 1998) or, at least, as an indication 
of different evolution properties of type 2 AGNs.
A different evolution of obscured objects may be linked to the
origin of the X--ray absorption as predicted
in scenarios where the absorption takes place in a starburst region
surrounding the nucleus (Fabian et al. 1998). 

While there is increasing evidence on the fact that AGNs supply the bulk 
of the XRB, the model(s) parameter space is still  
rather unexplored. In particular the type 2 objects evolution and XLF, 
the distribution of absorbing column densities and the 
AGNs high energy spectrum constitute a 
fundamental piece of information for all the models, which need to be tested 
by future observations. 

\begin{figure}
\centerline{\psfig{file=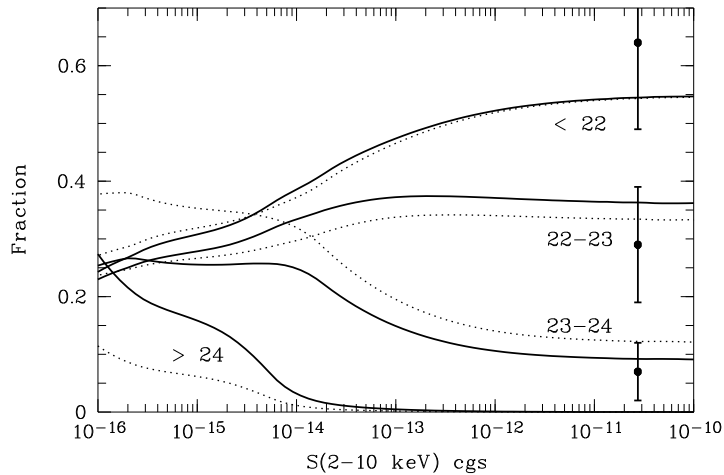, width=10cm, angle=-90}}
\caption{FIGURE 2. The relative fraction of objects with different absorption
column densities as a function of the 2--10 keV flux. 
The points at $\sim$ 3 $\cdot$ 10$^{-11}$ cgs 
represent the observed absorption distribution in the Piccinotti et al.
 (1982) sample.}
\end{figure}

\ssk
\ni
2.1  Testing the XRB models  
\ssk
\ni

The discovery of several heavily obscured, Compton thick ($N_H > 10^{24-25}$
cm$^{-2}$), objects in a BeppoSAX survey of 
a well defined sample of optically selected Seyfert 2 galaxies (Maiolino 
et al. 1998) suggests that the ``true" absorption distribution of type 2 AGN
might be different from that previously assumed (e.g. Madau et al. 1994, C95).
An unbiased estimate of the ``true" $N_H$ distribution of obscured 
objects could be obtained by X--ray spectroscopy of faint X--ray 
sources at high energies. As an example, the predictions of two 
self--consistent synthesis models (dotted and solid lines) following C95,
which differ only in the relative ratio of objects with different $N_H$ are 
reported in Fig. 2. 
The observed fractions of AGNs with different amounts of absorption 
start to deviate significantly from the local ones (derived from the Piccinotti 
et al. 1982 sample) only below $\sim$ 10$^{-13}$ cgs (2--10 keV).
Large area hard X--ray surveys reaching this limiting flux are currently
in progress (Giommi et al. 1998), while fainter X--ray fluxes will be 
reached in the near future by XMM and ABRIXAS allowing to probe
the $N_H$ distribution. 
The optical identification of these hard X--ray sources 
will open the opportunity to study the space density and evolution of type 2
objects and in particular to test whether type 2 QSOs exist. At present
the number of type 2 QSOs discovered so far (Georgantopoulos et al. 1998 
and references therein) is consistent with model predictions (Comastri 1998). 
Given that their space density is expected to rapidly increase at fainter 
fluxes, a relatively large number of type 2 QSO is expected in XMM and AXAF 
observations. 
Another assumption which needs to be tested by future observations 
concerns the broad band spectrum of AGNs. In order not to overproduce the
XRB intensity a break or an exponential cut--off in the high energy AGN spectra 
is required. The characteristic energy of this break is still poorly known.
Indeed the average Ginga/CGRO--OSSE spectrum of bright Seyfert galaxies 
in the $\sim$ 2--500 keV energy range 
suggests a cut--off at a few hundreds of keV (Gondek et al. 1996), also,
 with the exception of a few sources, the large majority of bright 
AGNs observed
by BeppoSAX (up to $\sim$ 200 keV) do not show evidence of a cut--off with 
lower limits of the order of 200--300 keV (Perola et al. 1998).
The results of two models which differ only in the
high energy cutoff are reported in Figure 3, suggesting a best fit value 
close to 500 keV.
Broad band high energy INTEGRAL observations will allow to address
this issue which is of the primary importance for the study of 
the emission mechanisms in AGNs.
Finally, the issue of the origin of the X--ray absorption and whether it 
originates from nuclear obscuration and/or from starburst driven activity 
as suggested by Fabian et al. (1998), may be tested exploiting the unique
spectral and spatial capabilities of XMM and AXAF respectively with 
simultaneous observations.

\begin{figure}
\centerline{\psfig{file=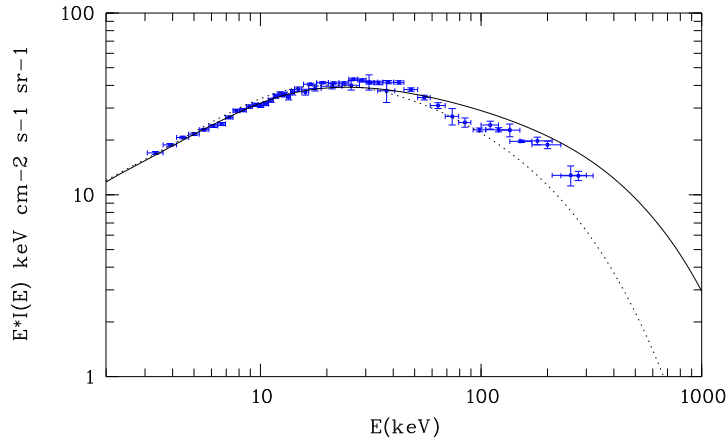, width=10cm, angle=-90}}
\caption{FIGURE 3. The effect of a different high--energy cut--off in the 
intrinsic AGN spectrum. Dotted line : $E_c$ = 300 keV , solid line : $E_c$ =
700 keV. }
\end{figure}

}

\ssk
\ni 3. AGN MODELS FOR THE GRB 
\ssk
\ni 

The unexpected discovery that a sub--class of AGNs, namely blazars, 
are strong $\gamma$--ray emitters was probably
the most important contribution of the CGRO EGRET instrument to the 
field of extragalactic astronomy. At present more than 50 sources have been 
clearly detected (Mukherjee et al. 1997). The large majority 
of them are flat spectrum radio--loud quasars (FSRQ),
while about a dozen are classified as BL Lacs. 
EGRET spectra (E $>$ 100 MeV), available for about half of the total sample, 
are well represented by a single power law with an average value
$\alpha \simeq$ 1.1--1.2.
There are some evidences of a departure from this simple shape,
possibly correlated with $\gamma$--ray intensity, as well as 
some differences between FSRQ and BL Lacs. 
These results are of great interest for the modeling of the GRB; 
however, we must await for more sensitive $\gamma$--ray observations of a 
larger sample of blazars to confirm the suggested trends.
The agreement between the average blazar spectrum and the GRB
strongly suggests that the bulk of the GRB is made by the summed 
contribution of unresolved blazars.
In order to compute the intensity of the GRB the knowledge of the
luminosity function and its evolution is needed.
One possibility is to extrapolate the radio luminosity function and
evolution in the $\gamma$--ray band using the observed correlation 
between radio and $\gamma$--ray luminosities (but see M\"ucke et al. 1997 
for a critical discussion on the robustness of such a correlation).  
Depending on the details of such extrapolation the contribution
of blazars to GRB ranges from 50~\% to 100~\% (Padovani et al. 1993, 
Setti \& Woltjer 1994, Stecker \& Salamon 1996).
In order to avoid errors introduced by the extrapolation one would have
to compute the $\gamma$--ray luminosity function
and evolution from the observed blazars sample. In this case, however, 
similar
uncertainties are introduced by the small number of $\gamma$--ray detected 
blazars.
Following this approach Chiang and Mukherjee (1998) conclude that only
about 25~\% of the GRB can be accounted for by blazars.
Similarly using a specific model for the blazars high energy emission  
M\"ucke and Pohl (1999) conclude that their contribution to the GRB is about 
40~\%. Given that the contribution of non--blazars sources, such as normal and
starburst galaxies is unlikely to exceed 25--30~\%  
(Soltan \& Juchniewicz 1999), a significant fraction of the GRB would 
remain unexplained. 
Two words of caution appear to be appropriate. First the 
true extragalactic intensity in the EGRET band may be lower 
due to some unaccounted Galactic contribution; second the estimate of
Chiang and Mukherjee (1998) is strongly dependent on the poorly known,
faint end slope of the luminosity function. More sensitive observations
are clearly needed in order to settle this issue. 
Whichever is the model of the high energy GRB, it appears that below 10 MeV 
the contribution of ``normal" blazars is even lower.
Indeed broad band (50 keV -- 1 GeV) $\gamma$--ray  observations 
(Mc Naron--Brown et al. 1995) of a small sample of sources indicate that
the average EGRET spectrum flattens significantly to a slope 
$\alpha \simeq$ 0.7 below a break energy of a few MeV.
Following the discovery by COMPTEL of a few FSRQ with a peculiar
X--$\gamma$--ray spectrum peaking at a few MeV (Bloemen et al. 1995),
it has been suggested that ``MeV blazars"
may provide a significant contribution to the COMPTEL background
(Bloemen et al. 1995; Comastri et al. 1996).
In figure 4 an AGN model for the full extragalactic background is reported.
The dashed line extending up to the highest energies represents the contribution 
of FSRQ to the GRB (of the order of 40~\% above 100 MeV), assuming a broken
power law spectrum with $\alpha_x$ = 0.5 below a break energy of 8 MeV and
 $\alpha_{\gamma}$ = 1.2 above. The evolution of the $\gamma$--ray 
luminosity function, derived from radio data, 
is that described in Comastri et al. (1996).
Assuming for the FSRQ belonging to the ``MeV blazars" class the same 
evolution of the $\gamma$--ray luminosity function of ``normal FSRQ", 
a broken power law
spectrum with  $\alpha_x$ = 0, $\alpha_{\gamma}$ = 2 and a break energy at 
1 MeV, their contribution to the GRB may be significant at energies 
between 100 keV and 1 MeV. The ``MeV blazars" contribution has been normalized
such that their 1 keV emissivity is about 15~\% of all FSRQ.  
This model is highly speculative, but it is interesting to note
that if this class of objects does indeed exist deep IBIS observations 
may reveal a few tens of them.

\begin{figure}
\centerline{\psfig{file=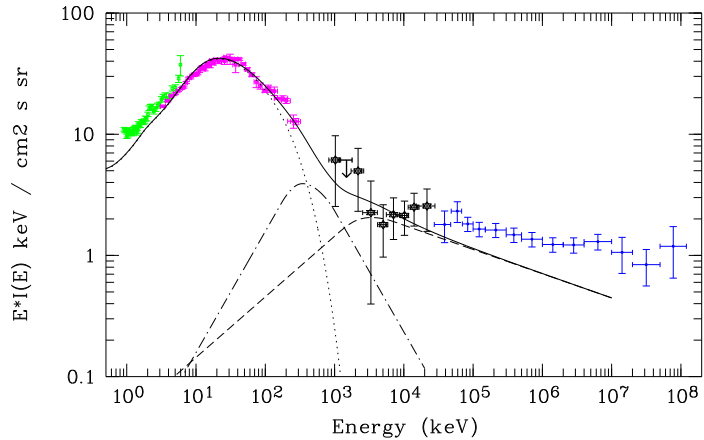, width=10cm, angle=-90}}
\caption{FIGURE 4. An AGN model for the XRB and GRB. Dotted line: XRB model 
of C95; dashed line: ``normal FSRQ" ; dot-dashed line: ``MeV blazars". See 
text for details.}
\end{figure}


\ssk
\ni 4. CONCLUSIONS 
\ssk
\ni 

On the basis of the present observations it is extremely likely that 
both the XRB and the GRB are mostly due to the summed contributions of 
different types of AGNs. The problem of the origin of the background(s)   
is now: which are the physical and evolutional properties of the 
sources (AGNs) making the background(s) ? 
The extremely faint fluxes which will be reached
by the next generation of X--ray missions (AXAF and XMM) will allow to study 
with unprecedented detail AGN evolution or, in other words, the history of 
accretion processes over the cosmic time. 
At high energies INTEGRAL will start to explore a new portion of the
spectrum filling the gap in the background measurements above a few hundreds 
of keV. INTEGRAL observations will also permit to investigate the 
high energy cut--off in AGN spectra providing an invaluable tool
to study the primary emission mechanisms. Finally INTEGRAL may uncover a 
class of sources, which may provide a significant contribution to the MeV 
background (i.e. the ``MeV blazars"), but have escaped detection in other 
bands. 

\ssk
\baselineskip = 12pt
{\abstract \ni ACKNOWLEDGMENTS

It is a pleasure to acknowledge G. Zamorani and G. Setti for useful 
discussions and G. Weidenspointner for providing some of the data in a computer
readable format.}

\ssk
\baselineskip = 12pt


{\references \ni REFERENCES
\ssk

\ref Bloemen H., et al., 1995, A\&A 293, L1

\ref Boyle B., et al., 1998, MNRAS 296, 1

\ref Chiang J., Mukherjee R., 1998, ApJ 496, 752 

\ref Cagnoni I., Della Ceca R., Maccacaro T., 1998, ApJ 493, 54

\ref Comastri A., Setti G., Zamorani G., Hasinger G., 1995, A\&A 296, 1

\ref Comastri A., Di Girolamo T., Setti G., 1996, A\&AS 120, 627

\ref Comastri A., 1998, Mem. S.A.It., in press (astro-ph/9809077) 

\ref Fabian A.C., Barcons X., Almaini O., Iwasawa K., 1998, MNRAS 297, L11

\ref Fichtel C.E., Simpson G.A., Thompson D.J., 1978, ApJ 222, 833

\ref Fiore F., et al., 1998, MNRAS, submitted

\ref Georgantopoulos I., et al., 1998, MNRAS in press (astro-ph/9810413)

\ref Giommi P., et al., 1998, Nuclear Physics B, 69/1--3, 591 

\ref Gondek D., et al., 1996, MNRAS 282, 646

\ref Gruber D.E., 1992, In The X--ray background, X. Barcons, A.C. Fabian 
(eds.), 44

\ref Halpern J. P., Eracleous M., Forster K., 1998, ApJ 501, 103

\ref Hasinger G. et al., 1998, A\&A 329, 482

\ref Madau P., Ghisellini G., Fabian A.C., 1994, MNRAS 270, L17

\ref Maiolino R., et al., 1998, A\&A 338, 781

\ref Matt G., Fabian A.C., 1994, MNRAS 267, 187

\ref Mc Naron--Brown K., et al., 1995, ApJ 451, 575 

\ref Miyaji T., Hasinger G., Schmidt M., 1999, Adv. Space Res., in press

\ref M\"ucke A., et al., 1997, A\&A 320, 33

\ref M\"ucke A., Pohl M., 1999, MNRAS, submitted

\ref Mukherjee R., et al., 1997, ApJ 490, 116 

\ref Padovani P., Ghisellini G., Fabian A.C., Celotti A., 1993, MNRAS 260, L21

\ref Perola G.C., et al., 1998, Mem. S.A.It., in press 

\ref Piccinotti G., Mushotzky R.F., Boldt E.A., et al., 1982, ApJ 253, 485

\ref Schmidt M., et al., 1998, A\&A 329, 495

\ref Setti G., Woltjer L., 1989, A\&A 224, L21

\ref Setti G., Woltjer L., 1994, ApJS 92, 629

\ref Soltan A.M., Juchniewicz J., 1999, Astrophys. Lett. and Comm., in press

\ref Sreekumar P., et al., 1998, ApJ 494, 523 

\ref Stecker F.W., Salamon M.H., 1996, ApJ 464, 600

\ref Ueda Y., et al., 1998, Nature 391, 866

\ref Watanabe K., et al., 1997, AIP 410 (2), 1223

\ref Weidenspointner G., et al., 1999, Astrophys. Lett. and Comm., in press

\ref Wright E.L., et al., 1994, ApJ 420, 450 
}                      

\end{document}